\documentclass[a4paper,pra,secnumarabic,preprintnumbers,twocolumn,amsmath,showpacs,showkeys,byrevtex]{revtex4}
\usepackage{graphicx}
\usepackage{amsfonts}
\newcommand{\mecanica}{\ensuremath{x}} 
\newcommand{\umax}[1]{#1_{\scriptscriptstyle{max}}}
\newcommand{\umin}[1]{#1_{\scriptscriptstyle{min}}}

\newcommand{\eref}[1]{(\ref{#1})}
\newcommand{\Eref}[1]{Equation~(\ref{#1})}
\newcommand{\fref}[1]{figure~\ref{#1}}
\newcommand{\sref}[1]{section~\ref{#1}}
\newcommand{\Sref}[1]{Section~\ref{#1}}
\newcommand{\Fref}[1]{Figure~\ref{#1}}

\begin{document}

\title[Universal restrictions to...]{Universal restrictions to the conversion of heat into work derived from the analysis of the Nernst theorem as a uniform limit}
\author{Jos\'{e} Mar\'{\i}a \surname{Mart\'{\i}n-Olalla}} 
\email{olalla@us.es}
\author{Alfredo \surname{Rey de Luna}}
\affiliation{Departamento de F\'{\i}sica de la Materia Condensada. Universidad de Sevilla. \\ Ap. Correos 1065 E-41080Sevilla. Spain}
\preprint{J.  Phys. A: Math. Gen. \textbf{36}(29) (2003) 7909--7921}

\begin{abstract}
The relationship between the Nernst theorem and the Kelvin-Planck statement of the second law is revisited. We put forward the fact that the exchange of entropy is uniformly vanishing as the temperature goes to zero. The analysis of this assumption shows that is equivalent to fact that the compensation of a Carnot engine scales with the absorbed heat so that the Nernst theorem should be embedded in the statement of the second law.
\end{abstract}

\maketitle

\section{Introduction}
\label{sec:intro}
The classical formulation of the Kelvin-Planck statement of the second law reads:\cite[page~89]{planck-1927}
\begin{quote}
``it is impossible to construct an engine which will work in a complete cycle, and produce no effect except the raising of a weight and the cooling of a heat-reservoir''
\end{quote}
Some other, essentially equivalent formulations of the law are possible ---Kelvin\cite{thomson-1853}, Clausius\cite{clausius-1854} and Carath\'eodory\cite{caratheodory-1909} statements--- but, for our purpose, we will refer to the formulation posed above. Essentially the statement requires  the presence of another  reservoir.

The development of the law needs the concept of  ``working fluid'', the substance that undergoes the cyclic process. The properties of the working fluid are usually discarded because the initial and final state of the fluid coincides and ``it has done  service  only as a transmitting agent in order to bring about the changes in the surroundings''\cite[page~68]{planck-1927}. Nonetheless, the fluid \emph{must} be able to do ``service'' in the way required.

We will show in this work that a general property of the matter will be restricting the ability of working fluids to perform cycles thus restricting what the  Kelvin-Planck statement allows. We will show that, in fact, such restriction follows from a comprehensive interpretation of the statement posed above.

The general property we are speaking about is nowadays known as the third law of thermodynamics. The necessity and character this law has been a matter of discussion from the early years of 1900's. Some chemical problems lead  Nernst\cite{nernst-1906,nernst-1924} to discover his \emph{heat theorem} which reads\cite[page~85]{nernst-1924}:
\begin{quote}
``in the neighborhood of the absolute zero \emph{all} processes proceed without alteration of entropy''
\end{quote}
The theorem ---which classically does not follow from the Kelvin-Planck statement\cite{einstein-1921,epstein-1937}--- is supported by a formidable array of experimental data. We choose this elder ---though very valuable--- version for the reasons that will be disclosed in \sref{sec:space} but we are seeing no particular reason for this statement to have been forgotten other than it refers to properties of processes rather than to properties of systems as it is nowadays stated\cite{kestin-1968ii,landau-lifshitz-1968,landsberg-78,baierlein-99,zemansky-1968}.

Nernst derived the theorem  from two quite general observations. The first one is the so-called \emph{principle of unattainability of the zero isotherm} which recalls the fact that no process can diminish the temperature of a system to the absolute zero. The second one is the fact that the specific heat of substances goes to zero as the temperature goes to zero.  

It should be acknowledge that Planck\cite{planck-1927} noticed that these observations should have lead to a ``more comprehensive'' conclusion: ``as the temperature diminishes indefinitely the entropy of a chemical homogeneous body of finite density approaches indefinitely to a definite value, which is independent of the pressure, the state of aggregation and of the special chemical modification.'' The Planck formulation avoids that $\Delta S\to0$ while $S\to-\infty$ as $T\to0$ and thus expresses that the absolute value of the entropy is bounded in the absolute zero.

Yet, our work will be just related to the analysis of $\Delta S$ since it just deals with the analysis of the conversion of heat into work. That problem is insensitive to a translation of the value of the entropy and thus, Planck's formulation lies out of our scope of interest. The same can be said about the vanishing of the specific heats.

Modern approaches and presentations of the third law of thermodynamics usually relates it to the microscopic properties of systems under consideration\cite{balian-91,huang-87,mafe-ajp-98,roseinnes-ajp-99,wu-pre-98}. Some efforts also try to clarify its macroscopic meaning\cite{falk-pr-1959,yan-chen-jpa-88,landsberg-jpa-89,oppenheim-jpa-89,liboff-physicsessays-94,landsberg-ajp-97,belgiorno-pra-03a}.

The goal of this paper is a revision of the mathematical description of the statement of the Nernst theorem posed above as well as its physical consequences. We will study pure macroscopic observations in the field of low temperature physics. In so doing no hypothesis about the constitution of the systems under study will be considered. 

Also, it is a burden of this manuscript an energetic analysis of the consequences of the statement posed above. To put it shortly, the simplest, most efficient engine contains two processes in which entropy is altered. This kind of processes are restricted at the neighborhood of absolute zero by the Nernst theorem; we will proof that such restriction will lead to a further condition  that any engine must satisfy.

\section{Limitations to the description of the Nernst theorem}
\label{sec:limit}

The classical formulation of the Nernst theorem reads:\cite{kestin-1968ii} ``the change in entropy associated with any isothermal process between two states of a system in internal equilibrium vanishes in the limit of zero temperature,'' which is usually translated into the mathematical condition:
\begin{equation}
\label{eq:weak}
    \forall\mecanica_1\mecanica_2\in\mathcal{D}\quad\lim_{T\to0^+}[S(T,\mecanica_1)-S(T,\mecanica_2)]=0
\end{equation}
where $\mecanica$ is any mechanical variable such as volume, pressure or magnetic field and $\mathcal{D}\subseteq\mathbb{R}$ is its domain of definition.\footnote{Nernst himself clothed \eref{eq:weak} through the statement posed in  \sref{sec:intro}.\cite[page~85]{nernst-1924}} 

Landau and Lifshitz\cite{landau-lifshitz-1968} showed the importance of keeping $\mecanica_1,\mecanica_2$ fixed in \eref{eq:weak}. Otherwise, they said, if, for instance, $\mecanica_1$ goes to infinity, the theorem may no be valid. Quite generally it could be said that that description works fine if the values of $\mecanica$ are indeed fixed but it comes into trouble in case one looks for double limits of the form $T\to0$ and $\mecanica\to\infty$.

\begin{figure}[tb]
  \centering
  \includegraphics[bb=131 406 465 720,scale=0.5]{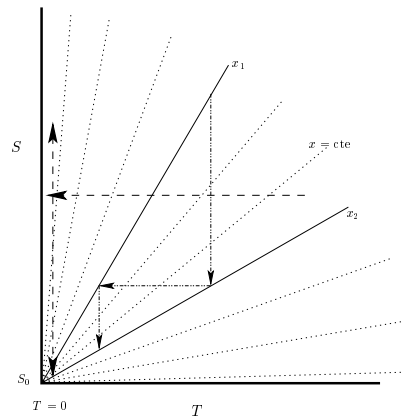}
  \caption{$T-S$ plot for the model \eref{eq:modelo}. Iso-$\mecanica$ lines are depicted. The unattainability statement and \eref{eq:weak} are satisfied for fixed $\mecanica_1$ and $\mecanica_2$. Yet an isentropic process ---horizontal left arrow--- diminishes the temperature of the system arbitrarily. Also any alteration of entropy is possible in the neighborhood of $T=0$ ---vertical double arrow--- provided that $\mecanica$ would change appropriately.}
  \label{fig:modelo}
\end{figure}

An academic example of this problem is provided by the following naive model
\begin{equation}
  \label{eq:modelo}
S(T,\mecanica)=S_0+\chi T\mecanica^g \Longrightarrow S(T,\mecanica_1)-S(T,\mecanica_2)=\chi T(\mecanica_1^g-\mecanica_2^g)\quad
\end{equation}
with $\mecanica\in\mathbb{R}^+$. In this expression $\chi$ is a positive constant that fits the dimensions of the model. Notice that  \eref{eq:modelo} satisfies \eref{eq:weak} but the double limit $T\to0,\mecanica\to\infty$ depends on the path it is achieved. \Fref{fig:modelo} depicts a $T-S$ plot for a system satisfying \eref{eq:modelo}.

The model does not fulfill the unattainability statement. First, it true that for fixed $\mecanica_1$ and $\mecanica_2$ the zero isotherm is unattainable\cite{zemansky-1968}; this proposition is essentially equivalent to \eref{eq:weak}. Yet, there is no need for so doing if one wants to achieve the neighborhood of the zero isotherm and it is also true that any isentropic path is endless for that model as ---let $g>0$--- by increasing indefinitely the mechanical parameter, the temperature is indefinitely decreasing to zero in a single step through $T\mecanica^g=\mathrm{cte}$. The ``unattainability''  of the zero isotherm would be a matter of practical limitations ---how to get an infinite $\mecanica$--- rather than a  fundamental restriction posed by a law of nature ---strictly speaking the absolute zero is here attained asymptotically.---  

Finally, the model does not accomplish for the words given by Nernst ---see \sref{sec:intro}--- either. No matter how close to zero the temperature can be \emph{any alteration} of entropy is possible provided that the mechanical variable increases sufficiently.

The ideal gas behaves similarly to that model. A classical description of the particles leads to an entropy with no lower bound and which does not accomplish for the Nernst theorem. On the contrary a quantum description of the problem leads to\cite{landau-lifshitz-1968,huang-87,wu-pre-98} \eref{eq:modelo} where $\mecanica$ is the volume and $g=2/3$ for fermions. A deep analysis of the model shows that the quantum or classical description is driven by the condition\cite{landau-lifshitz-1968,balian-91}:
$$
\frac{V}{N}\left(\frac{mkT}{2\pi\hbar^2}\right)^{3/2}\gg1
$$
known as ``classical limit.'' Here $N$ is the number of particles, $\hbar$ is Dirac's constant, $m$ is the particle mass and $k$ is the Boltzmann's constant. Thus, the competition between $T\to0$ and $V\to\infty$ is again crucial. It is likely that these contradictions are due to the fact that interactions are unavoidable at the very limit $T\to0$ so that microscopically ideal models may not be reflecting the array of data on macroscopic systems\cite{balian-91}.

We have shown then that a model satisfying \eref{eq:weak} ---the classical description of the Nernst theorem--- does not lead to the unattainability statement. Even worse, it does not fit to the words given by Nernst. Thus, further assumptions are required to a comprehensive and accurate mathematical description of the empirical laws observed at very low temperatures.

\section{The Nernst theorem as a uniform limit}
\label{sec:space}

The mathematical description of the statement of the Nernst theorem posed in Sec.~\ref{sec:intro} is improved by considering the following \emph{hypothesis I}: \emph{the isothermal exchange of entropy is \emph{uniformly} vanishing as the temperature goes to zero}:
\begin{equation}
  \label{eq:uniform}
  \forall\epsilon>0\quad\exists\delta(\epsilon)>0:\: T<\delta\Rightarrow \left|S(T,\mecanica_1)-S(T,\mecanica_2)\right|<\epsilon
\end{equation}
The key question\cite{jones-82} of the ``uniform convergence'' is that the \emph{same} $\delta(\epsilon)$ fits \emph{for any} $\mecanica_1,\mecanica_2$ belonging to $\mathcal{D}$.  It is straightforward that \eref{eq:uniform} matches the statement posed in \sref{sec:intro}. We will argue that it is the best choice to express mathematically that proposition.

The Nernst theorem is classically supposed to be  restricting the functional dependence of isothermal exchange of entropy $\Delta S=S(T,\mecanica_2)-S(T,\mecanica_1)$  on $T$ so that it converges to zero at the absolute zero. The ``uniform'' condition, here presented, essentially means that no value of $\mecanica$ can challenge this convergence. That is, no accidental divergence can possibly occur for a given value of $\mecanica$ in the neighborhood of $T=0$. In this way, the Nernst theorem would be also restricting the functional dependence of $\Delta S(T,\mecanica)$ on $\mecanica$.

\paragraph{R\^ole of $x$:}
\label{con:x}
If hypothesis I is taken into account, the mechanical variable plays no r\^ole in the description of the problem. That is the primary consequence of the uniform convergence since given $\epsilon$ then $\delta$ is just a property of the system under consideration regardless the value of $x$. The reader should notice that this is a burden in the formulation of the Nernst theorem as a law of nature which does not depends on the configuration of the system under consideration. 

In the classical description of the theorem \eref{eq:weak} $\delta$ is a function of $\epsilon,\mecanica_1,\mecanica_2$ so the mechanical variable does play a r\^ole in the description of the problem. Although this r\^ole is usually discarded it is of the most importance when considering, for instance, double limits.

\paragraph{Existence of inaccessible regions in a $T-S$ plot:}
\label{con:inaccessible}
\Eref{eq:uniform} ensures that $S(T,\mecanica_1)-S(T,\mecanica_2)$ is  bounded in the neighborhood of $T=0$ so that it has a supremum:
\begin{equation}
  \label{eq:sigma}
  \sigma(T)=\underset{\mecanica\in\mathcal{D}}{\sup}\big\{S(T,\mecanica_1)-S(T,\mecanica_2)\big\}
\end{equation}
The function $\sigma$ exists and is a monotonically increasing function at least in the neighbourhood of $T=0$. The function depends on the thermophysical properties of the system under consideration. What follows describe the relevant properties of the function in that neighbourhood. 

Now, consider a system compliant with hypothesis I whose equilibrium state is defined by a given temperature and a given mechanical configuration. The entropy of this state equals $S(T,\mecanica)$. Let us suppose that entropy is isothermally increased, the existence of $\sigma$ ensures that the final entropy cannot exceed $S(T,\mecanica)+\sigma$. The same argument applies for a process that decrease entropy. Hence, $S(T,\mecanica)$ is an upper and lower bounded function of $\mecanica$ for a given temperature and exists the following functions:
\begin{equation}
  \label{eq:sigma2}
  \umax{S}(T)=\underset{\mecanica\in\mathcal{D}}{\sup}\{S(T,\mecanica)\};\quad\umin{S}(T)=\underset{\mecanica\in\mathcal{D}}{\inf}\{S(T,\mecanica)\}
\end{equation}

Since the stability condition $(\partial S/\partial T)_\mecanica>0$ holds in the neighborhood of $T=0$ ---except perhaps at $T=0$--- the preceding functions are increasing functions of $T$ so that states of the type $\{T,S>\umax{S}(T)\}$ and $\{T,S<\umin{S}(T)\}$ cannot exist.  Hence, equilibrium states do not fill the plane $T-S$ and two boundaries arises from the fulfillment of the Nernst theorem as an uniform condition. 

\begin{figure}[tbp]
  \centering
\includegraphics[bb=143 381 492 726,scale=0.6]{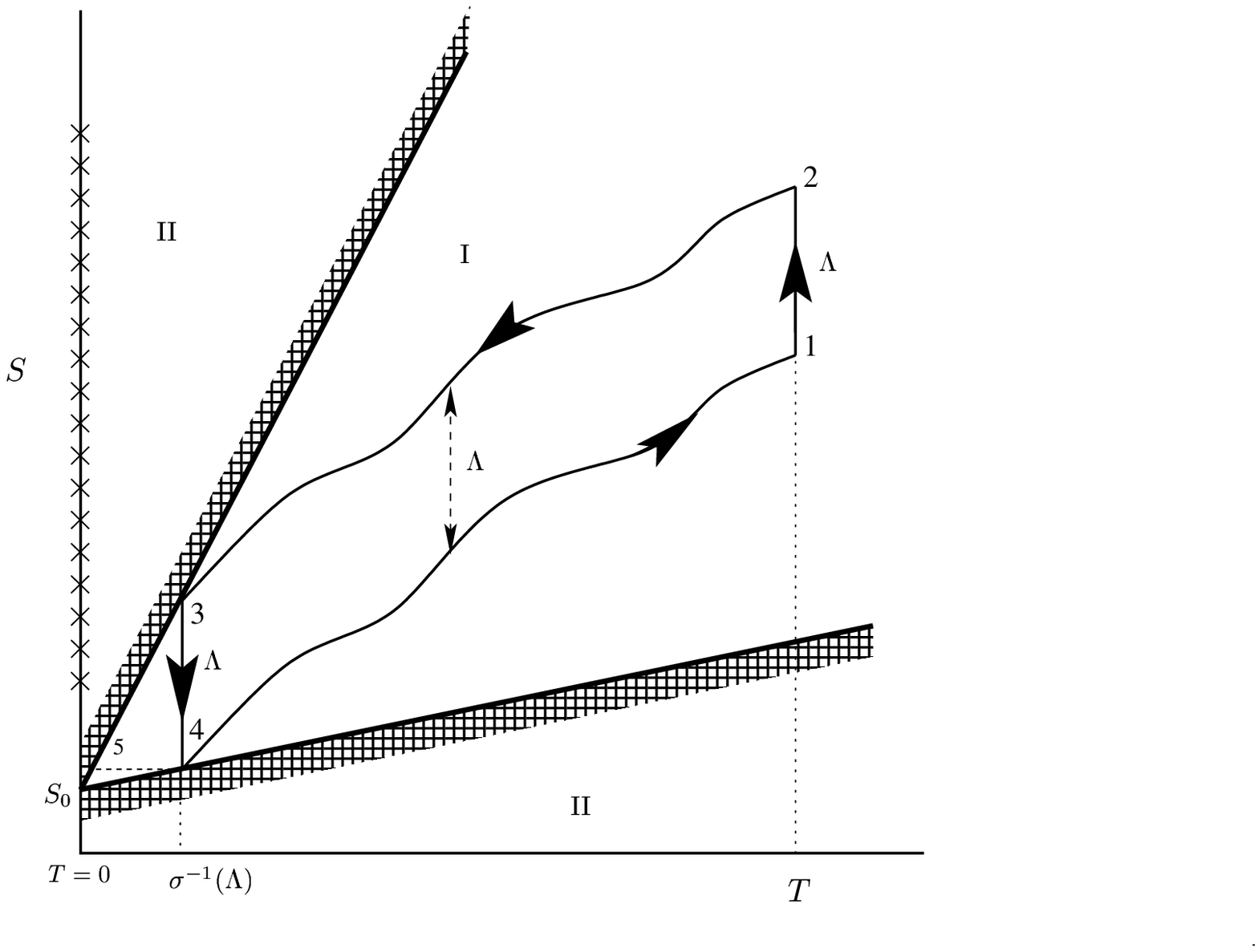}
\caption{The $T-S$ plot and the Nernst theorem. The symbol $\times$ represents the points classically excluded by the theorem\cite[Figure~23.5]{kestin-1968ii}. A more comprehensive analysis of the theorem reveals that there exist a region~I whose points  represents equilibrium states, and a region~II which does not do so. The cycle 1-2-3-4-1 is an engine consisting in two isotherms 1-2 and 3-4, and two processes, 2-3 and 4-1, which differ just in a shift of entropy $S'=S+\Lambda$. Area 1-2-3-4-1 is equal to $\Lambda[T-\sigma^{-1}(\Lambda)]$ as in a Carnot engine. The essential of this picture is valid even if $S_0\to-\infty$ while $\Delta S\to0$}
\label{fig:f}
\end{figure}

In more detail it could be said that one of the goals of the third law is to ensure that the entropy $S$ has a single value   in the neighborhood of zero isotherm\cite{kestin-1968ii}.  In the classical formulation of the theorem, points of the type $\{T=0,S\neq S_0\}$ are excluded\cite[Figure~23.5]{kestin-1968ii} in a $T-S$ plot (these points are sketched by the symbol $\times$ in \fref{fig:f}). From a physical point of view ``when a certain point is excluded, we must demand that the same must be true about a small region surrounding the point''\cite{caratheodory-1909}\cite[page~236]{kestin-1976}.\footnote{Carath\'edory is here speaking about the adiabatic inaccessibility and he is preparing his celebrated Axiom~II. It is noteworthy that the Axiom literally states that ``(\ldots) there exist states that cannot be approached \emph{arbitrarily close} by adiabatic process'' instead of bare phrase ``states that are \emph{inaccessible} by adiabatic process.'' The concept ``arbitrarily close'' or ``neighborhood'' will play a leading r\^ole in the following discussion.} On the contrary, if a neighborhood of $\{T=0,S\neq S_0\}$ could be reached,  the exclusion of those ---isolated--- points would be fictitious. 

Thus, the plot $T-S$ (see \fref{fig:f}) consists in the region~I of allowed values of $\{T,S\}$, and  the forbidden region~II. The existence of region II is a goal of hypothesis I. The boundaries  ---which may or may not be physically accessible--- does not coincide with the axis $T=0$. In \Fref{fig:f} and in the preceding discussion we have make use of the Planck hypothesis for the purpose of clarity. The same argument would apply if $S_0$ comes down to $-\infty$ and, simultaneously, $\Delta S$ is vanishing.

The following issue is known, however hypothesis I enriches and clarifies its meaning. 

\paragraph{Processes that come to an end (unattainability statement):}
\label{con:unattainability}

Let us consider the isentropic process $S=\Sigma_0$ starting at some temperature so that $\umin{S}<\Sigma_0<\umax{S}$. The process will go on until the temperature $T_1$ defined by $\umax{S}(T_1)=\Sigma_0$ is attained. That temperature is non zero.

Here we may decrease the entropy of the system isothermally until the condition $\Sigma_1=\Sigma_0-\sigma(T_1)=\umin{S}(T_1)$ is achieved. At that point, an isentropic process will cool the system down to the temperature $T_2$ defined by $\umax{S}(T_2)=\Sigma_1$.

The endless staircase process that goes to the absolute zero is then defined.

\paragraph{Vanishing of the thermal expansions coefficients}
\label{con:expansion}
 
The thermal coefficients are related to the derivative $(\partial S/\partial\mecanica)_T$ through Maxwell's relations\cite{kestin-1968ii}. From \eref{eq:uniform} it is derived the vanishing of such derivative since:
$$
\lim_{T\to0}\left(\frac{\partial S}{\partial\mecanica}\right)_T=\lim_{T\to0}\lim_{\mecanica'\to\mecanica}\frac{S(T,\mecanica')-S(T,\mecanica)}{\mecanica'-\mecanica}\quad\quad\forall\mecanica\in\mathcal{D}
$$
If this double limit exists, it can be computed in whichever order. By taking first $T\to0$ and invoking \eref{eq:uniform} one has, of necessity\cite{kestin-1968ii}:
\begin{equation}
  \label{eq:alfa}
  \lim_{T\to0}\left(\frac{\partial S}{\partial\mecanica}\right)_T=0\quad\quad\forall\mecanica\in\mathcal{D}
\end{equation}

Unlike properties a and b ---being mathematical propositions,--- it should be pointed out that properties c and d are trends\cite{nernst-1924,kestin-1968ii} amply confirmed by experiment providing a support for the hypothesis. Yet, the most important consequence of the hypothesis relates it to the problem of the conversion of heat into work and will be considered in detail in the following section.

\section{The Nernst theorem and the continuous production of work}
\label{sec:work}

In the preceding section we have shown that  the Nernst theorem forces the existence of forbidden regions in a $T-S$ plot. We now derive  consequences taking in mind that the uniformity condition introduced in the preceding section makes $\mecanica$ play no r\^ole in the problem.

Let us now consider the following question: we wish to build up an engine which produces mechanical work $W$ by using a given working fluid that draws a given  amount of heat $Q$ from a reservoir of a given temperature $T$, \emph{which is the minimal amount of heat $\umin{Q'}$ that is to be taken up to the cold reservoir?}\footnote{This question inspects the behaviour of $W$ ---or $Q'$--- once the hot reservoir and the absorbed heat are fixed. It is also customary to inspect the behaviour of $W$ once the hot and cold reservoirs are fixed. This problem is related to the concept of irreversibility\cite{landau-lifshitz-1968} and lies out of the scope of the following discussion.}

The classical reading of the Kelvin-Planck statement would just say that the heat taken up at the cold reservoir ---hereafter called ``compensation''--- must be non-zero $Q'\neq0$. It then seems that so long as this statement is concerned, a negligible $Q'$ would suffice. Thus, $\umin{Q'}=0^+$ ---ie arbitrarily close to zero but non-zero---. The answer is independent from $Q$, $T$ and the working fluid and comes from the fact we feel free to place a two-reservoir engine in a $T-S$ plot since no other restriction happens to be.

A machine having $\umin{Q'}=0^+$ would result in an efficiency $\eta=W/Q$ as close to unity as desired. This most efficient engine has never been built up; we will now put forward the fact that this is due to fundamental properties of matter despite of practical limitations to achieve such engine.

As a general rule, $\sigma$ is nonzero for nonzero temperatures and from \eref{eq:uniform} and \eref{eq:sigma} one gets:
\begin{equation}
  \label{eq:limsigma}
  \forall\epsilon>0\quad\exists\delta(\epsilon)>0:\quad T<\delta\Longrightarrow \sigma(T)<\epsilon
\end{equation}
That is, $\lim_{T\to0^+}\sigma(T)=0$.

In \eref{eq:limsigma} there is no need to call for absolute value delimiters since both $T$ and $\sigma$ are positive magnitudes. From \eref{eq:limsigma} and the preceding argument  the existence of the  inverse function $\sigma^{-1}(\Lambda)$ is straightforward. In fact, the inverse function is nothing else but a suitable representation for the parameter $\delta(\epsilon)$. The inverse function gives the temperature at which the width in entropy of the accessible states equals $\Lambda$. That temperature also depends on the thermophysical properties of the system under consideration.

Now, let us consider again the question posed at the beginning of this section but now consider that the Kelvin-Planck statement and the Nernst theorem, as stated in \sref{sec:intro}, apply. Hence, the restrictions posed in \sref{sec:space}, shown in \fref{fig:f}, are valid.  The working fluid is undergoing a cycle that is extracting an amount of entropy  $\Lambda=Q/T$ from the hot reservoir. For so doing, it is necessary that $\Lambda<\sigma(T)$.

Now, the entropy should be being deposited into  the cold reservoir which would receive an amount of heat $Q'$. To achieve the maximum efficiency, the temperature of the cold reservoir must be the coldest temperature able to exchange with the working fluid that  amount of entropy. Following the preceding paragraphs, that minimal temperature is given by $\sigma^{-1}(\Lambda)$ which is a property of the fluid under consideration. Thus,
\begin{eqnarray}
  \label{eq:eta}
  \forall Q\neq0,T\neq0:\, \frac{Q}{T}=\Lambda<\sigma(T)\nonumber\\
\Longrightarrow\exists \sigma^{-1}(\Lambda):\, Q'\geq\Lambda\times\sigma^{-1}(\Lambda)=\umin{Q'}
\end{eqnarray}
This most efficient engine is depicted in \fref{fig:f} by the cycle $1-2-3-4-1$ which consists of two isotherm and two processes that differ in a shift of entropy; exchanges of energy and entropy in $2-3$ cancel with those of $4-1$ so the reservoirs needed for these two processes play no r\^ole in the problem.\footnote{It would be possible to decrease $Q'$  by considering the cycle $1-2-3-5-4-1$ ---see \fref{fig:f}--- because entropy will be being deposited to colder reservoirs, but the condition of two reservoirs is broken. However, in that case, the compensation is equally expressed by $\sigma''\Lambda$ where $\sigma''$ is an unknown temperature ranging between $T_5$ and $T_4=\sigma^{-1}(Q/T)$. The essential of the following discussion also applies to this compensation.} The performance of work equals to  $\umax{W}=\Lambda\times\left[T-\sigma^{-1}(\Lambda)\right]$.

It is very noticeable that the value of $\umin{Q'}$ is now a function of the parameters of the problem: $Q$, $T$ and the working fluid which enters through $\sigma^{-1}$. Moreover, the minimal compensation is a function of the exchange of entropy $\Lambda$.

Though the particular value of $\umin{Q'}$ depends on the thermophysical properties of the working fluid it is a quite noteworthy fact that, as a general rule, $\umin{Q'}$ is never arbitrarily close to zero for a given $\Lambda$ since, according to the Nernst theorem, $\sigma^{-1}$ is not arbitrarily close to zero either (see \fref{fig:f}).

\begin{figure}[tbp]
  \centering 
 \includegraphics[bb=149 444 457 721,scale=0.6]{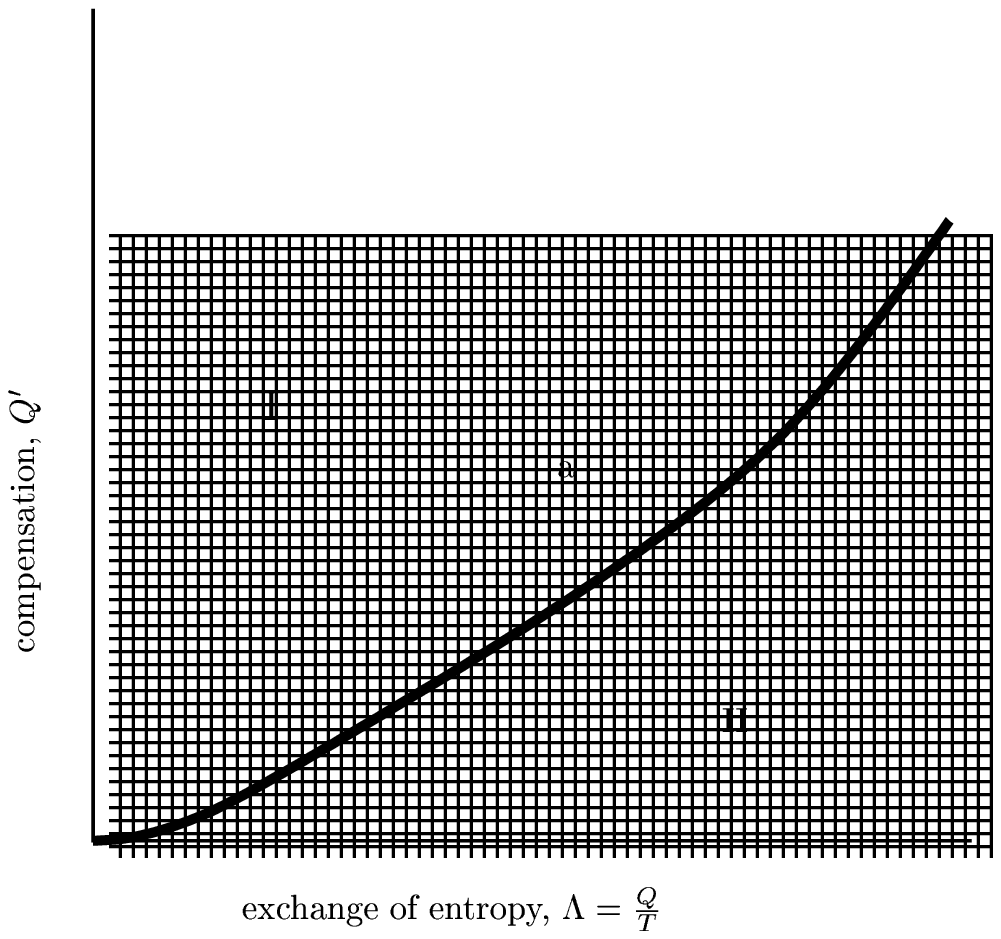}
  \caption{Plot of $\Lambda-Q'$ for a given working fluid. A point of region~I, say $a$, represents a set of engines each having the same $Q'$ and $\Lambda$ and the same temperature of the cold reservoir: the slope of the straight line.  The thick line ---which depends on the working fluid--- represents \eref{eq:eta} and comes to $\{0,0\}$ with zero slope. For the working fluid under consideration, it is impossible to built up an engine that enters in region~II.  The symbol $\times$ represents  the restriction posed by the classical reading of  the Kelvin-Planck. }
  \label{fig:etamax}
\end{figure}
The universal resemblance of \eref{eq:eta} allow to outline the plot $\Lambda-Q'$ (see \fref{fig:etamax}). Notice that in the context of the classical reading of the Kelvin-Planck statement that plot would have no restriction other than the exclusion of the points of the type $\{\Lambda\neq0,Q'=0\}$, those points are shown in that figure by the symbol $\times$. Now, if we take into account the Nernst theorem it is clear that points of the type $\{\Lambda,Q'<\umin{Q'}(\Lambda)\}$ should be excluded as well. That points define a region whose boundary is given by \eref{eq:eta}; the analogy between the regions I and II, and the boundary of \fref{fig:f} and those of \fref{fig:etamax} is immediate. From the Nernst theorem it is easily probed that $\umin{Q'}$ goes to zero with zero slope as $\Lambda$ goes to zero. Summarizing the meaning of \fref{fig:etamax} it should be stressed that  the Nernst theorem is excluding a region around those points actually excluded by the classical reading of the Kelvin-Planck statement. That is in agreement with Carath\'eodory's argument posed in \sref{sec:space}.

The limit $\Lambda\to0^+$ is the unique possibility to get $\umin{Q'}\to0^+$. In that case, taking $T$ as a bounded, constant parameter, it is clear that $Q$ will be vanishing as well and, as an ultimate consequence $W$ will be also vanishing. Hence:
  \begin{equation}
    \label{eq:qmin}
    Q'\to0^+\Longrightarrow W\to0^+
  \end{equation}
This proposition contains the essence of the Nernst theorem. Its dramatic meaning is best viewed noticing that $Q'\to0^+$ and $\eta\to1^-$ are equivalent. It then means that \emph{as the most efficient engine is achieved, the delivery of work is decreasing to zero}. 

It should be pointed out once again that in the preceding discussion $\mecanica$ does not play any r\^ole and it is in this sense that the restrictions \fref{fig:etamax} and \eref{eq:qmin} are \emph{universal}. That feature comes from the property of uniformity. On the contrary, if the Nernst theorem is just considered as a limit ---without the requirement of uniformity,--- then \eref{eq:qmin} and \fref{fig:etamax} would be just valid  for transitions between two given values of $\mecanica$ ---see \cite[Figure~23.9]{kestin-1968ii}\cite{liboff-physicsessays-94}.--- Therefore, \eref{eq:weak} does not lead to any proper restriction, as the boundary of \fref{fig:etamax} will come arbitrarily close to the restriction posed by the classical reading of the Kelvin-Planck statement provided that we consider the appropriate values for $\mecanica_1$ and $\mecanica_2$. Hence, the improvement of an engine would be a matter of practical limitations if \eref{eq:weak} were valid.\footnote{If the working fluid behaves like \eref{eq:modelo}, the engine could always be improved by getting arbitrarily large values of $\mecanica$.}

\section{The Nernst theorem and the statement of the second law}
\label{sec:second}
 
\Sref{sec:work} clearly shows that the Nernst theorem is restricting the conversion of heat into work in a way which is independent from the mechanical configuration of the system. It should be desirable to link it to the Kelvin-Planck statement of the second law.

One of the most important results of \sref{sec:work} is the leading r\^ole played by $Q'$, the compensation, in the problem of the conversion of heat into work. That importance comes from the fact that its minimum value is universally expressed by \eref{eq:eta}.  

\begin{figure}[tb]
  \centering  
  \includegraphics[bb=137 446 592 726,scale=0.5]{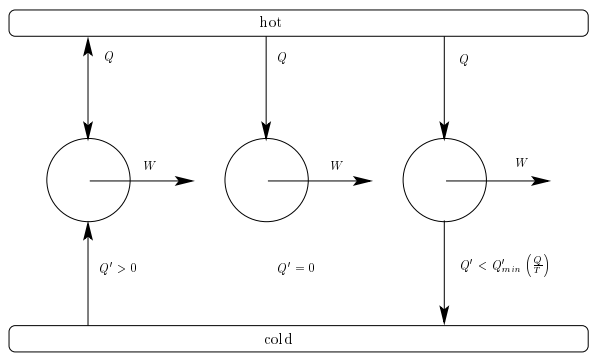}
  \caption{The staircase of the statements of the second law. The reader should notice the restrictions on the compensation that arises from the violations of the statements. From left to right a violation of the Kelvin statement, of the classical reading of the  Kelvin-Planck statement, and of a ``comprehensive'' reading of the Kelvin-Planck statement. In this last case, $\umin{Q'}$ is given by \eref{eq:eta}.}
  \label{fig:staircase}
\end{figure}

From an historical point of view, the r\^ole of the cold reservoir was  the key of the second law. The first known statement of the law, due to Kelvin\cite{thomson-1853}, stated that it is impossible to built up an engine that both produces work and  cools the coldest of the available reservoirs, no matter what happens to hotter reservoirs (see \fref{fig:staircase}).

Planck simplified the statement by noticing that it is impossible to built up an engine that performs work by cooling just one reservoir\cite{planck-1927}. In that sense, a cold reservoir is to be heated ---compensation--- in some amount (see \fref{fig:staircase}).  Yet he put forward no word about the ``size'' of the compensation so that one hopes it might  be negligible: that is the  ``classical'' reading of the statement that has prevailed from the beginning of the statement.

Furthermore, if the  Nernst theorem is taken into account through hypothesis I, we get the ultimate restriction: a minimal compensation is given by the properties of the working fluid and the exchange of entropy (see \eref{eq:eta} and \fref{fig:staircase}): \emph{wish you to transform a finite amount of heat into work, you must \emph{certainly} pay a  tax ---a compensation---, namely  the tax is not becoming finiteless at your willing}. The emphasized proposition is an informal statement for \eref{eq:qmin}.

The reader should notice that, in fact, \eref{eq:qmin}  is something else than the bare statement of the Nernst theorem since it also recalls the Kelvin-Planck statement. It is the dome that crowns the leitmotif of the principles of thermodynamics by putting forward the ultimate restriction on $Q'$. In fact, if the relation were clothed in a negative way the reader would have found a statement very much like any of the second law: \emph{it precludes delivering a \emph{finite} work without a \emph{finite} compensation}.

\section{From the Kelvin-Planck statement to the Nernst theorem}
\label{sec:inverse}

The preceding sections analyzed the Nernst theorem through the hypothesis I putting forward its close relation to the problem of the conversion of heat into work. Here we want to do the reverse trip: starting at the Kelvin-Planck statement, upon which assumptions is hypothesis I derived? 

In our opinion the classical reading of the Kelvin-Planck statement could be said to be ``crude'' in the sense that the cause ---heat absorbed from the hot reservoir--- and the unavoidable effect ---the minimal compensation--- has been being taking as uncoupled since the early stage of thermodynamics. A more comprehensive reading of the statement of the second law would have lead to some coupling between the $\umax{W}$ or $\umin{Q'}$ and $Q$ since a finiteless absorbed heat is necessary to get a finiteless compensation ---see \eref{eq:eta} and \fref{fig:f}.--- We put now forward the fact that this hypothesis ---labeled \emph{hypothesis II}--- suffice for the Nernst theorem:
\begin{quote}
  \emph{the compensation of a Carnot engine approaches indefinitely zero  \emph{only if} the heat absorbed from the reservoir is vanishing}
\end{quote}

The significance of the hypothesis would be revealed by the conclusions that it draws, however the reader should not conclude that hypothesis II is additional to the Kelvin-Planck statement since it is embedded in its words. The point is the meaning of the word ``effect'' that appears in the statement. One can get explicitly hypothesis II by changing ``effect'' by ``finite effect'' in the statement so that any ``finite'' absorbed heat necessarily leads to a ``finite'' compensation. But, in fact,  any effect is actually finite so that the modification would be a pleonasm.

The fact that the hypothesis leads to the Nernst theorem is surprisingly straightforward. Notice that the compensation of an engine equals $Q'=T'\times\Lambda$ where $T'$ is the temperature of the cold reservoir and $\Lambda$ the exchange of entropy.  That magnitude becomes zero (1) if $T'$ goes to zero regardless $\Lambda$ or (2) if $\Lambda$ goes to zero regardless $T'$. The hypothesis excludes option (1) that is, it excludes any alteration of entropy in the neighbourhood of the absolute zero. Hence, the Nernst theorem as stated in \sref{sec:intro} and as analyzed in \sref{sec:space} comes of necessity.

It is then concluded that hypothesis I and II are equivalent propositions so that the Nernst theorem ultimately follows from a comprehensive reading of  the Kelvin-Planck statement through hypothesis II.\footnote{Thus the second law ensures that the entropy goes to a value which does not depend on $\mecanica$ at the absolute zero. Its precise value, or whether this value is finite or infinite  ---Planck's formulation--- is alien to this description since the second law just concerns variations of entropy. } This proposition is valid  as far as  systems suitable to be working fluids of engines are being considered.

\section{Discussion and conclusion}
\label{sec:discu}

In 1909 Carath\'eodory\cite{caratheodory-1909}, following a suggestion by Born, successfully translated the classical statements of the second law ---which deal with the problem of production of work--- into a statement which refers to physical properties of an isolated system. This work introduces the reverse trip for the Nernst theorem. The theorem has been supposed to deal with the properties of systems in the neighborhood of $T\to0$; the study here  presented (see \sref{sec:work}, specially \eref{eq:eta} and \eref{eq:qmin})  relates the theorem back to the problem of production of work. The theorem will be expressing an universal property of the continuous production of work. 

The reader may ask which assumptions makes the Nernst theorem independent from the Kelvin-Planck and which does not do so. The goal of the second law of thermodynamics is to restrict the  continuous production of work putting forward the existence of a fundamental asymmetry: work is dissipated into heat but the reverse is not true. The goal enters by a statement which expresses, in words, a restriction.  We have shown in this paper that the degree of restriction matters and gets influence in the properties of systems. The classical reading of the Kelvin-Planck statement  assumes just that the compensation is nonzero. Upon this assumption the general properties of systems in the neighborhood of $T=0$ needs to be summarized as an independent law. Yet, a comprehensive reading of the statement through the cautious hypothesis II presented in \sref{sec:inverse} leads to some of these general properties: the unattainability statement and the vanishing of the expansion coefficients. We should again recall that the formulation here presented is insensitive to whether or not specific heats come to zero as the temperature comes to zero.

The new reading of the statement overcomes the embarrassing fact that $W$ must differ from $Q$ ---to what extend?--- by stating that they must do so in a measurable quantity which also depends on the working fluid which does play a r\^ole in the problem. The speech of Carath\'eodory quoted in \sref{sec:space} again makes the sense in this discussion: if the condition $W=Q$ is excluded, $W=Q^-$ should have been excluded as well.\footnote{We are presenting an analogy between the words by Carath\'eodory and the results of this work. It is not our willing to state that the former derives from the latter or conversely. It is a sort of coincidence that both problems speak about the same law of nature.} 

Any exception to the hypothesis would result in a failure of the consequences \emph{iii} and \emph{iv} quoted in \sref{sec:space} that is: (a) an experiment that would allow to increase $1/T$ indefinitely, (b) an experiment that would allow to decrease entropy indefinitely, or (c) a substance that would have non zero thermal expansion coefficients.\footnote{The reader should notice that the unattainability of the zero isotherm does not guarantee the Nernst theorem. The model $S(T,\mecanica)=\chi\log T\times (\mecanica+a)/(\mecanica+b)$ with $\chi,a\neq b>0,\mecanica\in\mathbb{R}^+$ provides a mathematical example of system which precludes experiments (a) and (b) but fails to obey the Nernst theorem.}

It is out of the scope of this paper to describe the microscopic relevance of the hypothesis, that is to determine which kind of Hamiltonians ---interactions--- would lead to uniformly vanishing $\Delta S$  as well as their symmetry properties. Well-known models ---specially ideal gases--- do not satisfy the hypothesis here presented. However the reader should not consider them as an ``exception'' to the hypothesis  since the hypothesis are grounded on experimental macroscopic basis and not on the analysis of microscopic models.

The relation between interactionless models and the Nernst theorem has been recently suggested\cite{balian-91,roseinnes-ajp-99} and it seems that the r\^ole of interactions can not be neglected in real systems at sufficiently low temperatures. The suggestion comes from the analysis of the independent spin system. That model does not satisfies the Nernst theorem since the ground state is degenerate. Yet, independent spin systems does not happen in nature since \emph{real} solids always exhibits magnetic correlation and ordering at sufficiently low temperatures. The resulting ordering ---usually a macroscopic new phase, either ferromagnetic or antiferromagnetic--- would satisfy the Nernst theorem.

Quite the same analysis  can be made on free particle systems. The reader should notice that the classical ideal gas allow to envision  a process of the type (a) and (b), moreover it satisfies (c) in contradiction with  the Nernst theorem. On its own,  quantum ideal gases\cite{landau-lifshitz-1968} gets vanishing thermal expansion coefficients and precludes experiment of the type (b) since entropy is necessarily a low bounded function. Yet, the quantum model still allows to envision type (a) experiments which do not happen to be in nature and which do preclude the meaning of the Nernst theorem. Of course in real systems ``ordering'' does always occur and the settlement of condensed phases seems unavoidable.

It is then likely that interactionless models ---either kinetic, magnetic or whichever--- do not accurately describe the properties of \emph{real} systems at sufficiently low temperatures since interactions can not be neglected.

\acknowledgements

The authors wish to thank to Mr. Jos\'e Antonio P\'erez G\'omez for long discussions and, specially, for enlightening them with the uniform convergence. The authors also thanks to Ms. Pilar N\'u\~nez for her kindly search of old bibliographic data.

This work was supported by Spanish \emph{Ministerio de Ciencia y Tecnolog\'\i a} project BFM2002-02237.

\bibliographystyle{apsrev}
\bibliography{libros,articulos}

\end{document}